\documentclass[twoside]{dis07}
\usepackage[latin1]{inputenc}
\usepackage[dvips]{graphicx,epsfig,color}
\usepackage{wrapfig,rotating}
\usepackage{amssymb,amsmath,array}

\begin{document}
\title{BFKL NLL phenomenology of forward jets at HERA and Mueller Navelet
jets at the Tevatron and the LHC}

\author{Christophe Royon
%
%
\vspace{.3cm}\\
%
DAPNIA/Service de physique des particules, \\ CEA/Saclay, 91191 
Gif-sur-Yvette cedex, France
%
}

\maketitle

\begin{abstract}
We perform a BFKL-NLL analysis of forward jet production at HERA which leads
to a good description of data over the full kinematical domain. We also
predict the azimuthal angle dependence of  
Mueller-Navelet jet production at the Tevatron and the LHC using the BFKL NLL
formalism.
\end{abstract}

\section{Forward jets at HERA}

\begin{wrapfigure}{r}{0.5\columnwidth}
\centerline{\includegraphics[width=0.45\columnwidth]{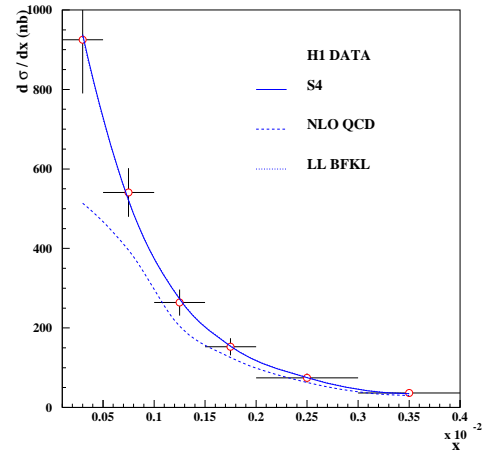}}
\caption{Comparison between the H1 $d \sigma /dx$ measurement 
with predictions for BFKL-LL, BFKL-NLL and DGLAP NLO calculations
(see text). S4 and LL BFKL cannot be distinguished on that figure.}\label{Fig1a}
\end{wrapfigure}

Following the successful BFKL \cite{bfkllo} parametrisation of the forward-jet 
cross-section $d\sigma/dx$ at Leading Order (LO)
at HERA \cite{mr,mrb}, it is possible to perform a similar study using Next-to-leading (NLL)
resummed BFKL kernels.
This method can be used for  
forward jet production at HERA in particular, provided one takes into account the proper symmetric {\it 
two-scale} 
feature of the forward-jet problem, whose scales are in this case $Q^2,$ for 
the lepton vertex and 
$k_T^2,$ for the jet vertex. In this short report, we will only discuss the
phenomelogical aspects and all detailed calculations can be found in Ref.
\cite{fwdjet} for forward jets at HERA and in Ref. \cite{mnjet} for Mueller Navelet jets
at the Tevatron and the LHC.

\subsection{BFKL NLL formalism}
We perform a saddle point
approximation of the BFKL NLL formalism and  compare it with the H1 forward
jet cross section measurements \footnote{We are in the process of checking
that implementing the full BFKL NLL kernel instead of performing a saddle point
approximation does not change the results of this paper and the quality of the
fits.}. 
The BFKL NLL \cite{bfkl} formalism reads:
\begin{eqnarray}
\frac{d \sigma}{dx} &=& N  \left( \frac{Q^2}{k_T^2} \right)^{\gamma} \alpha_S(k_T^2) 
\alpha_S(Q^2) ~ \sqrt{A} ~ \nonumber 
\exp \left(  \alpha_S(k_T Q)  \frac{N_C}{\pi} 
\chi_{eff}(\gamma_C)
\log (\frac{x_J}{x}) \right) \nonumber \\
&~&
\exp \left( - A \alpha_S(k_T Q) \log^2 (\sqrt{\frac{Q}{k_T}}) 
\right) \nonumber
\end{eqnarray}
with 
\begin{eqnarray}
A^{-1} &=& \frac{3 \alpha_S(k_T Q)}{4 \pi} \log \frac{x_J}{x} \chi_{eff}''
(\gamma_C) \nonumber \\
\gamma &=& \gamma_C + \frac{\alpha_S(k_T Q) \chi_{eff}(\gamma_C)}{2} \nonumber
\end{eqnarray}
where
the saddle point equation is  $\chi_{eff}'(\gamma_c)=0$. The effective kernels 
$\chi_{eff}(p,\gamma,\bar{\alpha})$ are obtained from the NLL kernel by solving the implicit equation:
\begin{eqnarray}
\chi_{eff}=\chi_{NLL}(p,\gamma,\bar{\alpha} 
\chi_{eff}) . \nonumber
\end{eqnarray} 
The values of $\chi$ are
taken at NLL \cite{bfkl}  using different resummation schemes to remove spurious
singularities defined as CCS, S3 and S4 \cite{resum}. Contrary to LL BFKL, it is
worth noticing that the coupling constant $\alpha_S$ is taken using the
renormalisation group equations, the only free parameter in the fit being the
normalisation.

\begin{figure}
\centerline{\includegraphics[width=0.75\columnwidth]{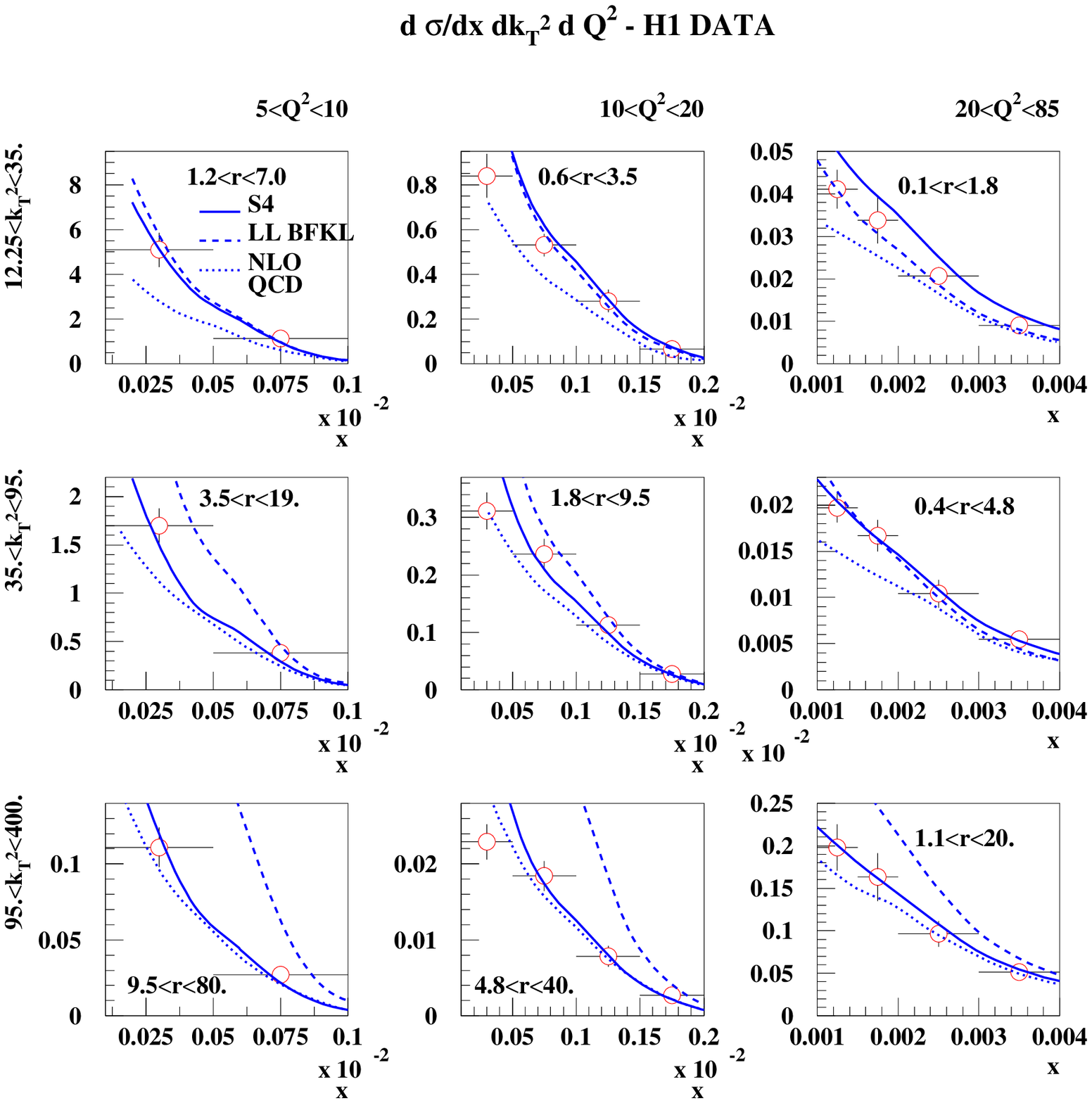}}
\caption{Comparison between the H1 measurement of the triple differential cross
section with predictions for BFKL-LL, BFKL-NLL and DGLAP NLO calculations
(see text).}\label{Fig1}
\end{figure}

One difficulty arises while fitting H1 $d \sigma/dx$ data \cite{h1} : we need to integrate
the differential cross section on the bin size in $Q^2$, $x_{J}$ (the momentum
fraction of the proton carried by the forward jet), $k_T$ (the jet transverse
momentum), while taking into account the experimental cuts. To avoid numerical
difficulties, we choose to perform the integration on the bin using the
variables where the cross section does not change rapidly, namely $k_T^2/Q^2$,
$\log 1/x_{J}$, and $1/Q^2$. Experimental cuts are treated directly at the
integral level (the cut on $0.5<k_T^2/Q^2<5$ for instance) or using a toy Monte
Carlo. 
More detail can be found about the fitting procedure 
in Appendix A of Ref. \cite{mrb}. 

The NLL fits \cite{fwdjet} can nicely
describe the H1 data \cite{h1} for the S4 scheme ($\chi^2=5.6/5$ per degree
of freedom with statistical errors only) 
whereas the S3 and CCS schemes show higher $\chi^2$.
($\chi^2=45.9/5$ and $\chi^2=20.4/5$ respectively with statistical errors only)
The fit $\chi^2$ are good for all schemes if one considers
statistical and systematics errors added in quadrature \cite{mr,mrb}.
The DGLAP NLO calculation fails to
describe the H1 data at lowest $x$ (see Fig. 1). 

The H1 collaboration also measured the forward jet triple differential cross
section \cite{h1} and the results are given
in Fig. 2. The BFKL LL formalism leads to a good description 
of the data when $r=k_T^2/Q^2$
is close to 1 and deviates from the data when $r$ is further away from 1. This
effect is expected since DGLAP radiation effects are supposed to occur when
the ratio between
the jet $k_T$ and the virtual photon $Q^2$ are further away from 1. The BFKL 
NLL calculation
including the $Q^2$ evolution via the renormalisation group equation leads to a
good description of the H1 data on the full range. We note that the higher order
corrections are small when $r \sim 1$, when the BFKL effects are
supposed to dominate. By contrast, they are significant as expected when $r$ is different from
one, ie when DGLAP evolution becomes relevant. We notice that the DGLAP NLO calculation
fails to describe the data when $r \sim 1$, or in the region where
BFKL resummation effects are expected to appear.

\section{Mueller Navelet jets at the Tevatron and the LHC}

\begin{wrapfigure}{r}{0.5\columnwidth}
\centerline{\includegraphics[width=0.45\columnwidth]{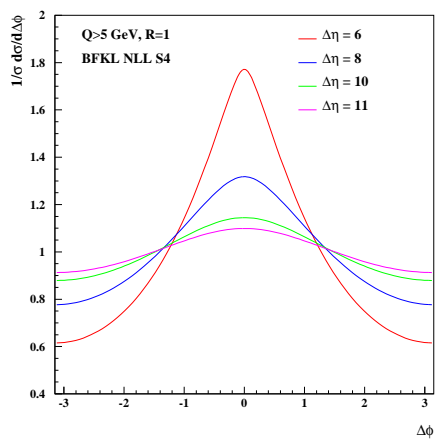}}
\caption{Azimuthal correlations between jets with $\Delta \eta=$6, 8, 10
and 11 and $p_T>5$ GeV in the CDF acceptance. This measurement will represent a
clear test of the BFKL regime.}\label{Fig1}
\end{wrapfigure}

Mueller Navelet jets are ideal processes to study BFKL resummation effects
\cite{mn}.
Two jets with a large interval in rapidity and with similar
tranverse momenta are considered. A typical observable to look for BFKL effects
is the measurement of the azimuthal correlations between both jets. The DGLAP
prediction is that this distribution should peak towards $\pi$ - ie jets
are back-to-bacl- whereas
multi-gluon emission via the BFKL mechanism leads to a smoother distribution.
The relevant variables to look for azimuthal correlations are the following:
\begin{eqnarray}
\Delta \eta &=& y_1 - y_2  \nonumber \\
y &=& (y_1 + y_2)/2 \nonumber \\
Q &=& \sqrt{k_1 k_2} \nonumber \\
R &=& k_2/k_1  \nonumber 
\end{eqnarray}
The azimuthal correlation for BFKL reads:
\begin{eqnarray}
2\pi\left.\frac{d\sigma}{d\Delta\eta dR d\Delta\Phi}
\right/\frac{d\sigma}{d\Delta\eta dR}=
1+ \nonumber 
\frac{2}{\sigma_0(\Delta\eta,R)}\sum_{p=1}^\infty \sigma_p(\Delta\eta,R) \cos(p\Delta\Phi)
\nonumber 
\end{eqnarray}
where in the NLL BFKL framework,
\begin{eqnarray}
\sigma_p&=& \int_{E_T}^\infty \frac{dQ}{Q^3}
\alpha_s(Q^2/R)\alpha_s(Q^2R) \nonumber 
\left( \int_{y_<}^{y_>} dy x_1 f_{eff}(x_1,Q^2/R)x_2f_{eff}(x_2,Q^2R) 
\right) \nonumber
\\
&~& \int_{1/2-\infty}^{1/2+\infty}\frac{d\gamma}{2i\pi}R^{-2\gamma}
\ e^{\bar\alpha(Q^2)\chi_{eff}(p, \gamma, \bar{\alpha})\Delta\eta}  \nonumber
\end{eqnarray}
and $\chi_{eff}$ is the effective resummed kernel.
Computing the different $\sigma_p$ at NLL for the resummation schemes S3 and S4
allowed us to compute the azimuthal correlations at NLL. As expected, the
$\Delta \Phi$ dependence is less flat than for BFKL LL and is closer to the
DGLAP behaviour \cite{mnjet}. To illustrate this result, we give in Fig. 3 the azimuthal
correlation in the CDF acceptance. The CDF collaboration installed the
mini-Plugs calorimeters aiming for rapidity gap selections in the very forward
regions and these detectors can be used to tag very forward jets. A measurement
of jet $p_T$ with these detectors would not be possible but their azimuthal
segmentation allows a $\phi$ measurement. In Fig. 2, we display the jet
azimuthal correlations for jets with a $p_T>5$ GeV and $\Delta \eta=$6, 8, 10
and 11. For $\Delta \eta=$11, we notice that the distribution is quite flat,
which would be a clear test of the BFKL prediction. Similar measurements are
possible at the LHC and predictions can be found in Ref. \cite{mnjet}.

\begin{footnotesize}


\end{footnotesize}


\end{document}